# Light Absorption Properties of the High Quality Linear Alkylbenzene for the JUNO Experiment

De-Wen Cao,[1] Rui Zhang,[1] You-Hang Liu,[1] Chang-Wei Loh,[1] Wei Wang,[1] Zhi-Qiang Qian,[1] Yu-Zhen Yang,[1] Xing Peng,[2] Yong-feng Zhang,[2] Ai-Zhong Huang,[3] Ming Qi,[1*]

[1] National Laboratory of Solid State Microstructures and School of Physics, Nanjing University, Nanjing 210093, China

[2] ThermoFisher Scientific, Shanghai 201206, China

[3] Jinling Petrochemical Corporation Ltd, Nanjing 210000, China

[*] Corresponding author: qming@nju.edu.cn

## Abstract

The Jiangmen Underground Neutrino Observatory (JUNO), a 20 kton multi-purpose underground liquid scintillator detector designed to determine the neutrino mass hierarchy, and measure the neutrino oscillation parameters. The excellent energy resolution and the large fiducial volume anticipated for the JUNO detector offer exciting opportunities for addressing many important topics in neutrino and astro-particle physics. Linear alkylbenzene (LAB) will be used as the solvent for the liquid scintillation system in the central detector of JUNO. The light attenuation lengths of LAB should be comparable to the diameter of the JUNO detector, hence very good optical transparency is required. However, the presence of impurities in the LAB renders an intrinsic limit for the transparency. This work focuses on the study of the effects of organic impurities in the LAB, and their light absorption properties particularly in the wavelength region of 350-450 nm. we have prepared LAB samples and measured their light attenuation lengths. These samples were then analyzed by a gas chromatography–mass spectrometry (GC-MS), and the structure formulas of organic impurities were ascertained. These impurities' light-absorption properties in the wavelength region of 350–550 nm were theoretically investigated with PCM-TDDFT. The overall optical transparency of the LAB samples was studied, which would further help us in promoting the LAB preparation technique for the mass production thereof, thus improving the transparency of the high quality LAB samples in the near future.

## 1. Introduction

The Jiangmen Underground Neutrino Observatory (JUNO) has been proposed to determine the neutrino mass hierarchy using an underground liquid scintillator detector[1-4]. It is located about 53 km away from both Yangjiang and Taishan Nuclear Power Plants in Guangdong, China. The experimental hall, spanning more than 50 meters, is under a granite mountain of over 700 m overburden. The JUNO is designed to consist of a 34 m diameter spherical antineutrino detector acting as the central detector (CD). The CD will contain 20,000 tons of liquid scintillator (LS) and 18500 photomultiplier tubes (PMT), and is designed to have a very good energy resolution of 3%. The LS in the detector serves as the target for the detection of the reactor antineutrinos. The current choice of the LS is: linear alkyl benzene (LAB) as the solvent, doped with PPO as the scintillation fluor and Bis-MSB as the wavelength shifter. A low-background environment is crucial since only a few tens of antineutrino events will be detected per day. As a consequence, light yield, fluorescence time profile, transparency and radio-purity are the key features of a good high quality LS.

For the emitted photons to reach the PMTs from their production vertex located within the detector, good optical transparency is required for the LAB. The optical transparency of the LAB is quantified by its attenuation length. The LAB samples for use in the JUNO experiment should achieve an attenuation length greater than 30 m, comparable to the diameter of the spherical detector, If the LAB does not reach a very high level of transparency, then the flickering light from the neutrino signal is likely to be absorbed by the liquid itself and could not reach the PMTs. The preparation of a highly transparent liquid scintillator poses a huge challenge to the experiment.

Industrial produced LAB, for example as a raw material in the production of detergents, can contain a lot of impurities, which could reduce the transparency of the LAB. We used gas chromatography technique (Q Exactive GC-MS, Thermo Fisher) with a high resolution to analyze the impurities of from LAB samples manufactured from the Nanjing LAB plant. The results show that some of the impurity molecular groups with characteristic absorption wavelengths are ubiquitous in LAB samples. Further calculations and analysis show that although several impurities are on the order of several ppb in concentrations in the LAB, they could still affect the light transmission quality of the LAB, resulting in a decrease in the attenuation length of the LAB. As a comparison, the Daya Bay experiment [5-8] with a 5-meter diameter and height cylindrical detector contains LAB with a light attenuation length of only 10 m. The researchers achieved the goal of 10 m attenuation length by identifying and removing several impurities of relatively high concentration [9-10]. Since our goal for this research work is a light attenuation length of nearly 30 m, the challenge to prepare such LAB is much greater for the JUNO detector.

## 2. LAB Samples Preparation

LAB was proposed as a liquid scintillator solvent in JUNO for its attractive properties, including rich H atoms content, appreciable light yield, high flash point (130 $^{o}$C, which could significantly reduce safety concerns for the JUNO experiment), relatively low cost, non-toxicity and thereby environmentally friendly. Apart from JUNO and Daya Bay, LAB has also been widely used in SNO+ [11] and RENO [12], as well as the veto detector of Double Chooz [13]. Compared with other detectors, the LAB in the JUNO experiment needs a higher attenuation length for light with wavelengths within the range of 350 - 550 nm, in particular at 430 nm. Regrettably, such high quality LAB is not available from commercial-based productions of LAB.

The LAB samples used in this work, tagged as NJ32#, NJ33# and NJ44#, were prepared with various methods. The NJ32# sample is a highly-purified sample prepared with aluminum oxide filtration technique at the manufacturing site of Jinling [14]. NJ33# is an initial sample prepared using an improved technique which is more suitable for large-scale mass productions in the future. NJ44# is a recent sample using the improved technique. These three samples were studied by measuring the attenuation lengths through a 1.2 m vertical tube-LED system. Then we used a gas chromatography technique (Q Exactive GC-MS, Thermo Fisher) with a high resolution to analyze their impurities through their light absorption spectrum.

## 3. Experiment

### 3.1 Attenuation length of LAB

By definition, the attenuation length $L_\lambda$ is the distance into a material when the intensity of an incident light with wavelength λ has dropped to 1/e. It can be expressed as follow[15]:

$$I = I_0 e^{\frac{-x}{L_\lambda}} \quad (1),$$

where I is the light intensity after it passes through the material, which in this work will be the LAB liquid, for a total path length x, and $I_0$ is the initial intensity of incident light.

The experimental setup of the measurement of the attenuation length was designed at the Nanjing University. It is composed of two parts; the first part being the optical measuring and controlling system, and the second part being the data acquisition system (DAQ) [16]. Through continuous improvement and upgrades, the systematic error of this system has been minimized effectively [17]. The measuring results of the three samples are shown in Table 1.

| LAB sample | Attenuation length at 430 nm |
|---|---|
| NJ32# | 26.8±0.4m |
| NJ33# | 16.9±0.3m |
| NJ44# | 25.8±0.7m |

Table 1. Attenuation lengths of NJ32#, NJ33# and NJ44#

**3.2 Composition analysis**

Gas chromatography–mass spectrometry (GC-MS) is an analytical method that combines the features of gas-chromatography and mass spectrometry to identify different substances within a test sample. The coupling of Orbitrap mass spectrometry with GC marks an exciting advance in the capability for GC/MS, offering a significant improvement in the resolving power, mass accuracy, sensitivity and linear range. Figure 1 shows the schematic diagram of Exactive GC-MS. We used GC-MS technique (Q Exactive GC-MS, Thermo Fisher) to analyze the LAB samples. This instrument has a high resolution (1 ng/L), and an accurate mass measurement across a wide

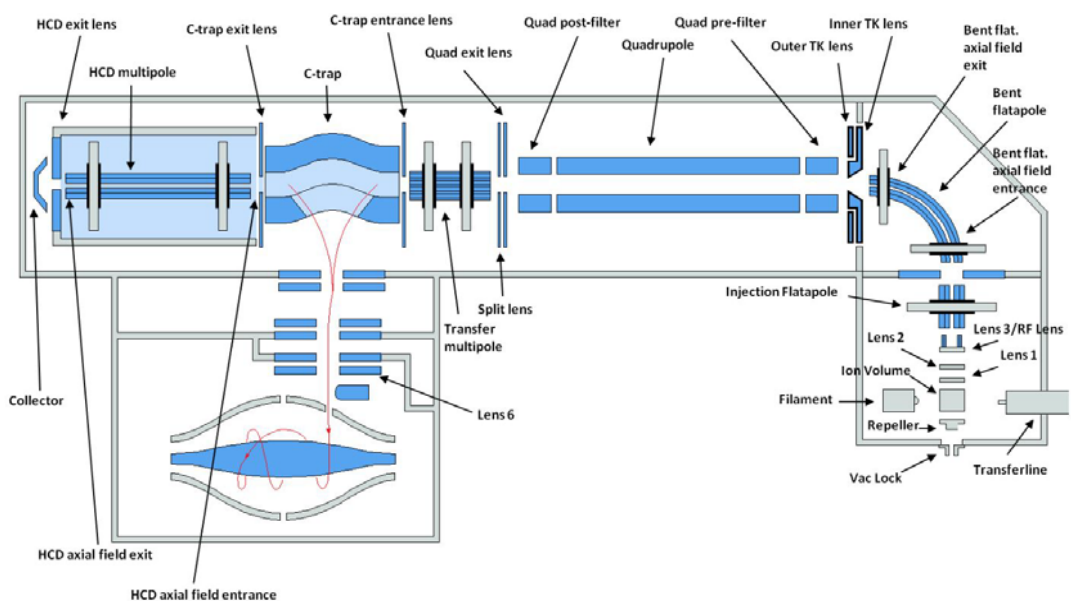

Figure 1: The schematic diagram of Exactive GC-MS

dynamic range, good linearity and the ability to identify impurities through Electron Ionisation (EI)

and Chemical Ionisation (CI) processes, which enables this instrument to be a powerful tool for impurity analysis. Figure 2 shows the GC-MS spectrum of the LAB sample, the 19 obvious peaks are the principal components of LAB with the chemical formula $C_6H_5+C_nH_{2n+1}$, but they only have light absorption peaks lower than 280 nm. The attenuation length of LAB samples in 350-550 nm is lower could be attributed to the non-alkylbenzene impurities in the LAB samples; it is the smaller peaks which are those that we need to pay more attention. We used GC-MS with a high resolution to make further analysis, various impurities were inspected, and the structure formulas were ascertained.

Theoretically, the absorption of UV-Vis optical wavelength concerns mainly with the electron transitions from the ground state to the excited state. Figure 3 shows the four kind of electron transitions corresponding to different absorption types of the optical wavelength in the organic compound, i.e. (1) $\sigma \rightarrow \sigma^*$, (2) $\pi \rightarrow \pi^*$, (3) $n \rightarrow \sigma^*$, (4) $n \rightarrow \pi^*$, respectively[18]. The red-shifted $\pi \rightarrow \pi^*$ and $n \rightarrow \pi^*$ transitions have a light absorption tendency at an UV-Vis optical wavelength region of 350-450 nm. The major $n \rightarrow \pi^*$ functional groups (chromophores) typically appear on the compound which contain benzene rings with nitrogen, sulfur, oxygen and chlorine [19]. We therefore paid more attention to those impurities. Table 2 shows some of the suspicious impurities which may greatly absorb light within our characteristic window (350nm - 450nm) in the NJ32#, NJ33# and NJ44# samples.

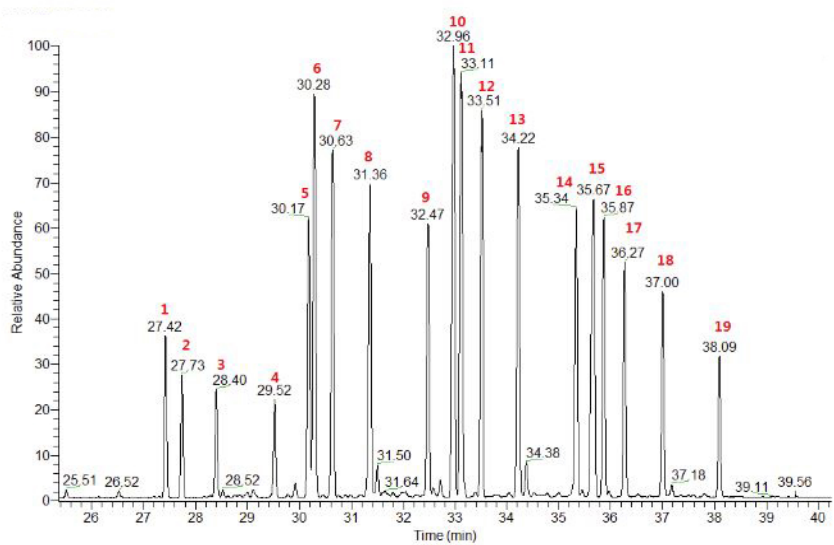

Figure 2: The GC-MS spectrum of the LAB sample

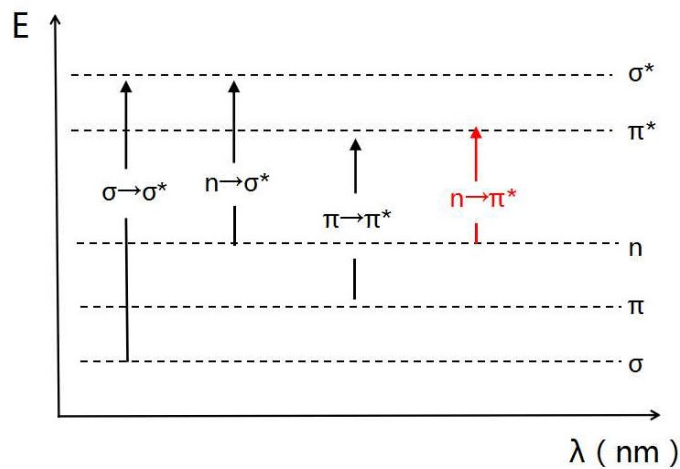

Figure 3: Types of electron transitions

| NJ32#(26.8m) | NJ33#(16.9m) | NJ44#(25.8m) |
|---|---|---|
| | C20H21N3O3S | |
| | C22H17NO6S | |
| | C23H26N2O2S | |
| | C25H31N5O | |
| | **C29H20N2O4** | |
| | C30H37N3O3 | |
| | **C33H38ClNOS** | |
| C18H16N2O4 | C18H16N2O4 | C18H16N2O4 |
| C25H28N4O2 | C25H28N4O2 | C25H28N4O2 |
| C25H29NO2 | C25H29NO2 | C25H29NO2 |
| C29H34N2O5 | C29H34N2O5 | C29H34N2O5 |
| C33H35NO11 | C33H35NO11 | C33H35NO11 |
| C33H42N2O7 | C33H42N2O7 | C33H42N2O7 |
| **C34H36N2O4** | **C34H36N2O4** | **C34H36N2O4** |

Table 2. Some of the impurities compounds in NJ32#, NJ33# and NJ44# samples.

## 3.3 Impurity molecules and calculation

The time-dependent density functional theory (TD-DFT) is well known as a powerful tool for ab initio quantum chemical studies of molecular optics and spectroscopy. TD-DFT combines the advantages of density functional theory and time-dependent formalism allowing the accurate determination of excited-state properties. For some special types of organic compounds, TD-DFT with suitable hybrid functional and extended basis set can get highly accurate data of molecules' lowest-energy light absorption wavelengths. For some special type compounds, TD-DFT with hybrid functional PBE0 [20] and large basis sets can get these special type compounds' low lying transition energy with a very high precision (within ±10nm). Denis Jacquemin [21] have studied more than 100 different thiocarbonyls and found that IEF-PCM-TD-PBE0/6-311+g (2df, p) // PBE0/6-311g (2df, p) is an excellent tool to predict the thiocarbonyl salts' properties of the lowest lying absorption. So we used the same method to investigate the light absorption properties of the impurity compounds in Table 1（the impurities compounds marked in blue exist only in the NJ33# sample, and the impurities compounds marked in green exist in all three samples）.

The impurity compounds' light absorption properties were calculated with the Gaussian 09 suite of programs by IEF-PCM-TD-PBE0/6-311+g (2df, p)//PBE0/6-311g (2df, p). To calculate the absorption peak wavelength(transition energy), we have used a three-step methodology: (1) an optimization of the ground state structure; (2) an analytic determination of the vibration spectrum in order to make sure no imaginary frequencies exists; (3) the evaluation of the electronic excited states.

## 4. Results and Discussion

Calculation results of suspicious impurities compounds are listed in Table 3. Three of the impurities were found to have absorptions in the 350 – 550 nm region, i.e. C29H20N2O4, C33H38ClNOS and C34H36N2O4 respectively (C29H20N2O4 is tagged as impurity I, C33H38ClNOS is tagged as impurity II, and C34H36N2O4 is tagged as impurity III), and are marked in bold.

| NJ32#(26.8m) | NJ33#(16.9m) | NJ44#(25.8m) | λmax (f) |
|---|---|---|---|
| | C20H21N3O3S | | 263.24 nm (0.0166) |
| | C22H17NO6S | | 342.15 nm (0.0782) |
| | C23H26N2O2S | | 251.56 nm (0.0298) |
| | C25H31N5O | | 312.07 nm (0.1564) |
| | **C29H20N2O4 (impurity I)** | | **440.80 nm (0.0268)** <br> **408.81 nm (0.0126)** <br> **374.05 nm (0.2698)** |
| | C30H37N3O3 | | 248.66 nm (0.0326) |
| | **C33H38ClNOS (impurity II)** | | **413.51 nm (0.0090)** |
| C18H16N2O4 | C18H16N2O4 | C18H16N2O4 | 292.85 nm (0.0139) |
| C25H28N4O2 | C25H28N4O2 | C25H28N4O2 | 292.49 nm (0.0071) |
| C25H29NO2 | C25H29NO2 | C25H29NO2 | 259.48 nm (0.0044) |
| C29H34N2O5 | C29H34N2O5 | C29H34N2O5 | 297.22 nm (0.0005) |
| C33H35NO11 | C33H35NO11 | C33H35NO11 | 368.24 nm (0.0400) |
| C33H42N2O7 | C33H42N2O7 | C33H42N2O7 | 261.34 nm (0.2124) |
| **C34H36N2O4 (impurity III)** | **C34H36N2O4 (impurity III)** | **C34H36N2O4 (impurity III)** | **505.93 nm (0.0203)** <br> **419.88 nm (0.1417)** <br> **419.88 nm (0.1417)** |

Table 3. PCM-TDDFT results of the impurities

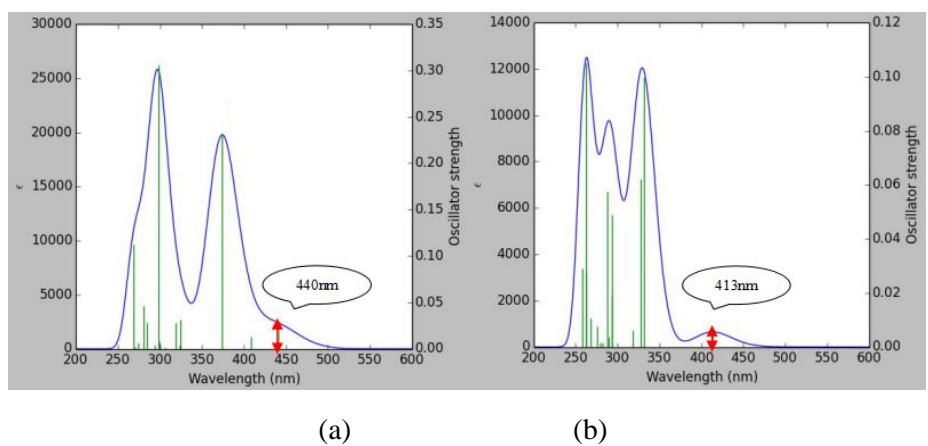

(a)  (b)

Figure 4 (a) and (b): The UV-Visible spectrometry of impurity I (C29H20N2O4) and impurity II (C33H38ClNOS).

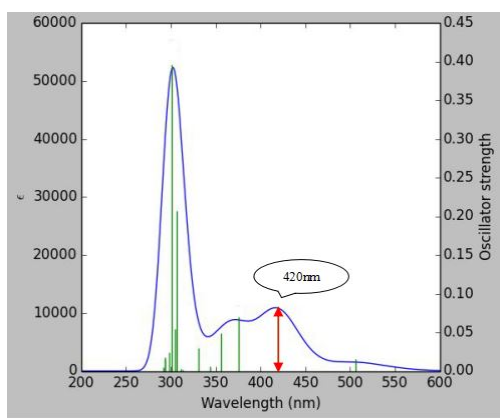

Figure 5: The UV-Visible spectrometry of impurity III (C34H36N2O4)

Considering the quantum calculation errors (within ±10nm), we find an agreement between the calculated absorption peaks and experimental UV-Vis absorption peaks. The three impurities could be the reasons for the low attenuation length in our characteristic wavelength window. Impurity I and Impurity II exist only in the NJ33# sample which has the lowest attenuation length among the samples. Figure 4 shows that they have absorption peaks around 440nm and 413nm respectively, which clearly decreases the attenuation length of the LAB sample. In Figure 5, impurity III which exist in all three samples was also found to have a strong absorption in the wavelength region of 350 - 450 nm. Table 4 shows the relative concentration ratio of impurity III and the attenuation length in three LAB samples, where the concentration in the three samples are normalized to the concentration in NJ32#. As can be observed from the Table 4, there is a negative correlation between the concentration of impurity III and the attenuation length of the sample. This suggests that, in order to achieve an attenuation length of more than that of NJ44#, the impurity III in the LAB samples should be reduced, if not eliminated.

| Sample | NJ32# | NJ33# | NJ44# |
|---|---|---|---|
| Attenuation length (m) | 26.8 | 16.9 | 25.8 |
| Relative concentration | 1 | 3.1 | 1.2 |

Table 4. Relative concentration ratio of impurity III and the attenuation length in three LAB samples

## 5. Summary

In this work, the light absorption properties of different LAB samples have been studied in details, particularly in the wavelength region of 350 - 450 nm. Three organic impurity compounds were detected in the LAB samples. They were studied with thorough on their light absorption properties with quantum calculations. The results show that they have characteristic absorptions in the wavelength region of 350 - 450 nm. Furthermore, their presence and relative concentration affect the attenuation length of the LAB sample, suggesting that the elimination of them could be the key in achieving our goal of at least 30 m in attenuation length.

## 5. Competing Interests

The authors declare that they there are no competing interests regarding the publication of this paper.

## Acknowledgments

We are grateful for the warm help and enlightening insights from Wang Yi-Fang, Cao Jun, Qian Sen, Zhou Li, Ding Ya-yun, Ning Zhe, Zhu Na, Yu Guang-You and Wang Wei. This work was supported by the National 973 Project Foundation of the Ministry of Science and Technology of China (Contract No. 2013CB834300).

[17] H. Yang, D. Cao, Z. Qian et al. *Light attenuation length of high quality linear alkyl benzene as liquid scintillator solvent for the JUNO experiment.* Journal of Instrumentation, Volume 12, November 2017 .

[18] R.M. Silverstein, G.C. Bassler and T.C. Morrill. *Spectroscopic Determination of Organic Compounds, 4th edition, John Wiley & Sons*, New York U.S.A. (1981).

[19] Griffiths J (1976) Colour and constitution of organic molecules. Academic Press, London

[20] J. P. Perdew, K. Burke, and M. Ernzerhof. *Generalized Gradient Approximation Made Simple*, Phys. Rev. Lett. **77**(1996) 3865.

[21] Jacquemin D, et al. *Ab Initio Investigation of the n → π* Transitions in Thiocarbonyl Dyes*, J. Phys. Chem **A 110**(2006) 9145.